# N-type doping of LPCVD-grown β-Ga$_2$O$_3$ thin films using solid-source germanium

Praneeth Ranga[1], Arkka Bhattacharyya[1], Luisa Whittaker-Brooks[2], Michael A. Scarpulla[1,3] and Sriram Krishnamoorthy[1]

[1]Department of Electrical and Computer Engineering, The University of Utah, Salt Lake City, UT 84112, United States of America

[2]Department of Chemistry, The University of Utah, Salt Lake City, UT, 84112, United States of America

[3]Department of Materials Science and Engineering, The University of Utah, Salt Lake City, UT, 84112, United States of America

a) Electronic mail: praneeth.ranga@utah.edu and sriram.krishnamoorthy@utah.edu

## Abstract

We report on the growth and characterization of Ge-doped β-Ga$_2$O$_3$ thin films using a solid germanium source. β-Ga$_2$O$_3$ thin films were grown using a low-pressure chemical vapor deposition (LPCVD) reactor with either an oxygen or gallium delivery tube. Films were grown on 6˚ offcut sapphire and (010) β-Ga$_2$O$_3$ substrates with growth rates between 0.5 – 22 μm/hr. By controlling the germanium vapor pressure, a wide range of Hall carrier concentrations between $10^{17} – 10^{19}$ cm$^{-3}$ were achieved. Low-temperature Hall data revealed a difference in donor incorporation depending on the reactor configuration. At low growth rates, germanium occupied a single donor energy level between 8 – 10 meV. At higher growth rates, germanium doping predominantly results in a deeper donor energy level at 85 meV. This work shows the effect of reactor design and growth regime on the kinetics of impurity incorporation. Studying donor incorporation in β-Ga$_2$O$_3$ is important for the design of high-power electronic devices.

# I.  INTRODUCTION

Power electronics based on UWBG (Ultrawide bandgap semiconductors) can lead to significant improvement in cost, performance and energy savings[1]. Recently, β-Ga$_2$O$_3$ has emerged as an attractive candidate for next-generation power electronics and deep-UV applications. β-Ga$_2$O$_3$ has a bandgap of 4.8 eV, which leads to a predicted breakdown field of 6-8 MV/cm, which would be significantly larger than other commercial power semiconductors[2]. The high breakdown field results in a very large Baliga's figure of merit (BFOM) which is ~ 3500x higher than for traditional semiconductor material such as silicon. A larger value of BFOM signifies smaller on-state conduction losses, which results in improved on/off state device performance. In addition to large BFOM, β-Ga$_2$O$_3$ has unique features not seen in other UWBG materials such as availability of large-area single crystal bulk substrates and orders of magnitude controllable n-type conductivity[3]. Significant progress has been made since the first realization of β-Ga$_2$O$_3$ MESFETs (metal semiconductor field effect transistors) in 2012[2]. This includes advances in growth[4–7], fabrication[8,9] and understanding fundamental properties of β-Ga$_2$O$_3$[10,11]. Lateral and vertical devices with critical breakdown field exceeding SiC and GaN have been demonstrated experimentally[12,13]. Devices with power densities reaching as high as 1 GW/cm$^2$ [14] and breakdown voltages up to 8 kV[8] were already realized. In addition to unipolar devices, rapid progress has been made in studying β-(Al$_x$Ga$_{1-x}$)$_2$O$_3$/β-Ga$_2$O$_3$ heterostructures[15–18] and integration of p-type materials with β-Ga$_2$O$_3$[19].

Based on its intrinsic material properties, β-Ga$_2$O$_3$ based power devices have the potential to reach breakdown voltages as high as 100 kV[20]. For reaching such high breakdown voltages, it is important to achieve high-quality, low-doped thick β-Ga$_2$O$_3$ drift layers. Choosing the correct epitaxial technique is critical to maximizing the performance of β-Ga$_2$O$_3$ power devices. The key criteria for choosing the proper thin film growth technique are growth rate, cost, material quality and unintentional impurity incorporation. Growth of β-Ga$_2$O$_3$ has been realized using a variety of techniques such as MBE (Molecular beam epitaxy)[21], MOCVD (Metal-organic chemical vapor deposition)[5–7,22,23], HVPE (Hydride vapor phase epitaxy)[4,24], PLD (Pulsed laser deposition)[25] and LPCVD (Low-pressure chemical vapor deposition)[26–29]. Growth of β-Ga$_2$O$_3$ has already been studied using MBE for a variety of film orientations (010), (-201), (001) and (110) and dopants(Si, Sn and Ge)[30–34]. A variety of lateral FETs have been fabricated using MBE-grown β-Ga$_2$O$_3$ films[2,18]. Whereas PLD is primarily utilized for studying heteroepitaxial growth of Ga$_2$O$_3$ polymorphs[35] and achieving heavily doped n$^+$ β-Ga$_2$O$_3$ layers[25]. Although high-quality material can be grown using most of the techniques, MBE and PLD are not suitable for the growth of thick drift layers. Only chemical vapor-based deposition techniques can offer growth rates reaching tens of μm/hr. In this regard, HVPE and MOVPE are the most promising and widely used methods for studying β-Ga$_2$O$_3$ growth. High-quality β-Ga$_2$O$_3$ films with mobility values close to the theoretical maximum have already been realized using HVPE and MOCVD. Since the start, HVPE is the most studied technique for the growth of thick, low-doped drift layers. HVPE growth of β-Ga$_2$O$_3$ is performed using GaCl$_3$ and O$_2$ as precursors and SiCl$_4$ as dopant gas[36]. By growing at high temperatures and large precursor flows, growth rates up to 250 μm/hr have been achieved[37]. However, MOVPE is preferred for studying β-(Al$_x$Ga$_{1-x}$)$_2$O$_3$/β-Ga$_2$O$_3$ heterostructures[16,17], since growth of β-(Al$_x$Ga$_{1-x}$)$_2$O$_3$ is not feasible using HVPE. In addition, the HVPE growth of high-quality β-Ga$_2$O$_3$ is largely limited to growth on (001) substrates[4]. On the other hand, high-quality films with record high mobility were realized in MOVPE-grown (010) and (100) β-Ga$_2$O$_3$ thin films[5–7,23]. Doping densities up to $10^{13}$ cm$^{-3}$ are attained using HVPE by compensating the background donor impurities. Because of using chlorine based precursor, a large amount of unwanted impurities have been found in as-grown HVPE β-Ga$_2$O$_3$ films[36].

In addition to other CVD-based techniques such as MOVPE and HVPE, LPCVD has emerged as a promising growth technique for studying the growth of β-Ga$_2$O$_3$. LPCVD is a low-pressure growth technique using elemental gallium and oxygen as precursors. Although MOVPE and HVPE are promising for the growth of β-Ga$_2$O$_3$, they use an expensive process and generate harmful byproducts. Unlike HVPE and MOVPE, LPCVD precursors are inert and reaction products are non-toxic. This leads to lower maintenance costs and simpler a deposition process. Since LPCVD uses elemental gallium instead of TEGa (Triethyl Gallium), the presence of external impurities could be potentially lower in LPCVD-grown β-Ga$_2$O$_3$ thin films. LPCVD growth of β-Ga$_2$O$_3$ has been studied using (010), (001) β-Ga$_2$O$_3$[38] and c-plane sapphire substrates[26,27,29]. Depending on the configuration, growth rates up to 15 μm/hr were achieved on offcut sapphire wafers[26]. Typically, LPCVD growths are performed in a mass transport limited regime. In this window, the growth rate dependence on substrate orientation is generally low. High-quality β-Ga$_2$O$_3$ films have been realized on a variety of growth orientations such as (010), (001) and (-201)[27,38]. Thin films grown on c-plane sapphire contain a large amount of rotational domains, which lead to poor electrical properties. By growing on a sapphire wafer with an intentional 6˚ offcut, high-quality (-201) β-Ga$_2$O$_3$ films were also realized[27]. Room temperature mobility values exceeding 100 cm$^2$/Vs were observed in both Si-doped homoepitaxial (010) and heteroepitaxial (-201) grown thin films[27,39]. By growing at high temperatures (950 ˚C - 1050˚C), room temperature Hall mobility of 150 cm$^2$/Vs was realized for a doping density of 1.5 x 10$^{17}$cm$^{-3}$ [39]. The measured Hall mobility values are very close to the highest reported mobility from HVPE and MOVPE thin films. Preliminary devices based on LPVCVD grown β-Ga$_2$O$_3$ thin films show breakdown fields up to 5.4 MV/cm, which is very encouraging for the realization of high-power β-Ga$_2$O$_3$ devices[40]. In spite of the huge promise, LPCVD growth studies of β-Ga$_2$O$_3$ are still limited. Most studies use elemental gallium and O$_2$ gas as growth precursors. SiCl$_4$ has been used as the silicon dopant. Studying the effect of using different n-type dopants and studying dopant incorporation as a function of growth conditions are important to realize high-quality β-Ga$_2$O$_3$ thin films.

In this work, we demonstrate n-type doping of LPCVD-grown films using solid germanium source. We developed a new technique for doping LPCVD-grown β-Ga$_2$O$_3$ thin films. Instead of using a gas-based dopant source, we utilized solid germanium as an n-type dopant precursor. By controlling the growth conditions, n-type doping is achieved for a range of doping concentrations. The carrier concentration, mobility and activation energy are extracted from temperature-dependent Hall measurements. Using two different growth reactor configurations, the donor incorporation is studied as a function of growth conditions and reactor configuration. The proposed process can be extended to study different kinds of donor and acceptor impurities in β-$Ga_2O_3$.

## II. EXPERIMENTAL

Growth of β-$Ga_2O_3$ is performed using a custom-built LPCVD reactor as shown in Fig.1. The setup consists of single-zone furnace with a quartz tube reactor. Elemental gallium (7N) is used as the gallium precursor. The Ga source is placed inside a quartz boat, which is positioned at the center of the reactor. Given the affinity for the oxidation of gallium, the Ga and $O_2$ vapors need to be separated ideally until they reach the substrate. Two different approaches have been utilized to minimize the amount of parasitic gas-phase pre-reactions. The first approach consists of supplying oxygen gas through a special delivery tube that physically isolates Ga and $O_2$ molecules (setup 1) until they reach the substrate surface (Fig. 1(a)). Argon gas is also separately injected to facilitate gallium transport. In setup 2, the gallium source is placed inside a 1- inch quartz tube which is placed inside the larger reactor (Fig. 1(b)). The argon push gas was connected to the 1-inch tube for controlling Ga transport. The oxygen gas is supplied through a separate line in order to reduce the amount of prereaction.

In general, the unintentionally-doped (UID) as-grown films were not conductive. An external dopant is required to induce n-type conductivity. Instead of using a gaseous dopant precursor, we used a solid source impurity for n-type doping. Given the low vapor pressure of Ge compared to Ga at 900 ˚C, solid Ge pellets were used as the Ge source. The amount of doping in the film is controlled by changing the Ge temperature and the amount of source material in the boat. Both offcut sapphire and (010) Fe-doped β-$Ga_2O_3$ substrates are used for the growth of β-$Ga_2O_3$. The samples are cleaned with acetone, IPA (iso-propyl alcohol) followed by a deionized (DI) water rinse. The samples are placed on a quartz stage which is then placed a few centimeters from the quartz tube at the center of the furnace.

The typical parameters utilized for β-Ga$_2$O$_3$ are growth temperature 800 ˚C – 900 ˚C, O$_2$ – 3 - 5 sccm, Argon – 70 - 140 sccm, Pressure – 0.5 - 2 Torr and growth time 0.5 – 2 hours.

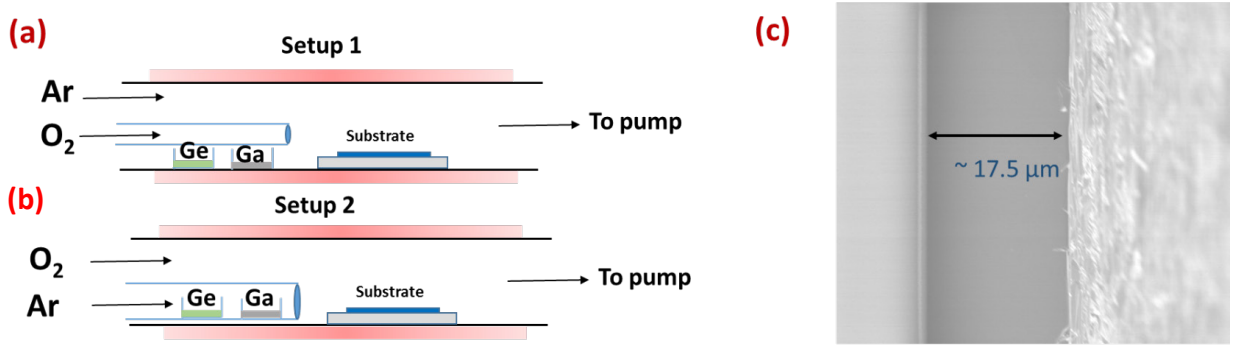

Fig.1 Schematic of the custom-built Ga$_2$O$_3$ LPCVD reactor with two different configurations- (a) Setup 1 (b) Setup 2. (c) Cross-sectional SEM image of the sample – grown using setup 2, showing a nominal thickness of 17.5 μm.

The thickness of the as-grown film was measured using cross-sectional scanning electron microscopy (SEM). Room temperature and Hall measurements were used to characterize carrier concentration, mobility of Ge-doped β-Ga$_2$O$_3$. Ti/Au (50 nm/50nm) Ohmic contacts were deposited by DC sputtering on the four corners of the sample to form a van der Pauw contacts. Low-temperature Hall measurements were performed for measuring low-temperature Hall mobility and donor activation energy. Ni/Au (50 nm/50 nm) layers are deposited for obtaining Schottky contacts to the epilayer. CV (capacitance-voltage) measurements were used to characterize the electrically active donor concentration of the doped thin films.

There are many previous reports on the presence of an unintentional second donor in β-Ga$_2$O$_3$[5,7,20,22]. We adopted a model of two donor levels to describe our temperature-dependent Hall data. The model assumes the presence of two donor energy levels and an ionized acceptor. The charge neutrality relation relating ionized donors, electron concentration and ionized acceptor concentration is shown in Eq.1[5,20].

$$n + N_a = \frac{N_{d1}}{1 + 2\exp\left(-\frac{E_{d1} - E_F}{k_B T}\right)} + \frac{N_{d2}}{1 + 2\exp\left(-\frac{E_{d2} - E_F}{k_B T}\right)} \qquad \text{Eq (1)}$$

Where $N_{d1}, N_{d2}$ are the densities of the two donor energy levels and $N_a$ is acceptor concentration. ($E_{d1}, E_{d2}$) and $E_a$ are the donor and acceptor energy levels respectively. $E_f$ is the Fermi level and n is the carrier concentration. $k_B$ is the Boltzmann constant and T is the measurement temperature. The Fermi level separation is calculated using the relation $n = N_c e^{-\frac{E_c - E_f}{k_B T}}$, where $N_c$ is the conduction band effective density of states calculated by assuming $m^* = 0.28 m_0$. The donor energy levels $E_{d1}$ and $E_{d2}$ are obtained by fitting the temperature-dependent carrier density data from 80 – 340 K for each sample.

## III. RESULTS AND DISCUSSION

First, we focus on Ge-doped β-Ga$_2$O$_3$ grown using setup 1. The typical growth temperature is in the range of 800 ˚C – 900 ˚C. The growth rate in this temperature range is limited by mass transport of the gallium and oxygen precursors. Typical growth rates in this configuration is in the range of 0.5 – 5 μm/hr. All the growths were done on c-plane sapphire substrates with 6˚ offcut towards a-plane. Based on the literature, using a c-plane sapphire wafer with an intentional offcut, suppresses the formation of rotational domains, which leads to the improved material quality of β-Ga$_2$O$_3$ films[27].

Fig.2(a) shows temperature-dependent Hall data for films grown on 6˚ offcut sapphire substrate. By adjusting the Ge source boat position Hall carrier concentrations between 5 x 10$^{17}$ cm$^{-3}$ – 1 x 10$^{19}$ cm$^{-3}$ were realized (samples A -D). A control sample was prepared for which no Ge dopant is used. Hall measurements revealed that the undoped sample is electrically insulating. This correlation indicates that Ge is the primary source of n-type conductivity in the LPCVD grown films. The room temperature mobility reduces from 25 – 5 cm$^2$/Vs with increasing carrier concentration. The reduction in mobility with increasing carrier concentration is attributed to higher ionized impurity scattering in heavily-doped films. Similar trends have been observed in epitaxially grown β-Ga$_2$O$_3$. In general, the low mobility values are attributed to the high density of defects due to the heteroepitaxial nature of the growth. Further growth optimization is required to improve electron mobility. Temperature-dependent Hall measurements indicated an Arrhenius relationship between carrier concentration and temperature. By fitting the low-doped films to the charge neutrality equations, activation energies between 8 – 10 meV are obtained. The Hall data could be completely explained by the presence of a single shallow energy level. Films with a lower amount of doping (Samples A and B), showed Arrhenius behavior as seen in typical doped semiconductors. In samples with a high amount of doping (C and D), the carrier concentration did not freeze out at low temperatures, indicating degenerate doping. This observation is attributed to impurity band formation usually observed in heavily-doped semiconductors. Similar behavior is observed in heavily Sn-doped substrates when doping exceeded the mott critical carrier density in gallium oxide ($n_{mott}$ ~ 2 x $10^{18}$ $cm^{-3}$)[41]. The low-temperature Hall mobility is observed to peak at 120 K, this is due to reduced ionized impurity scattering as the measurement temperature reduces. The peak mobility of low-doped films is still comparatively lower than existing literature. This is attributed to the presence of structural defects in the heteroepitaxial β-$Ga_2O_3$ epilayer. For samples with very high doping (sample D), the Hall mobility does not show a significant variation across the measurement range.

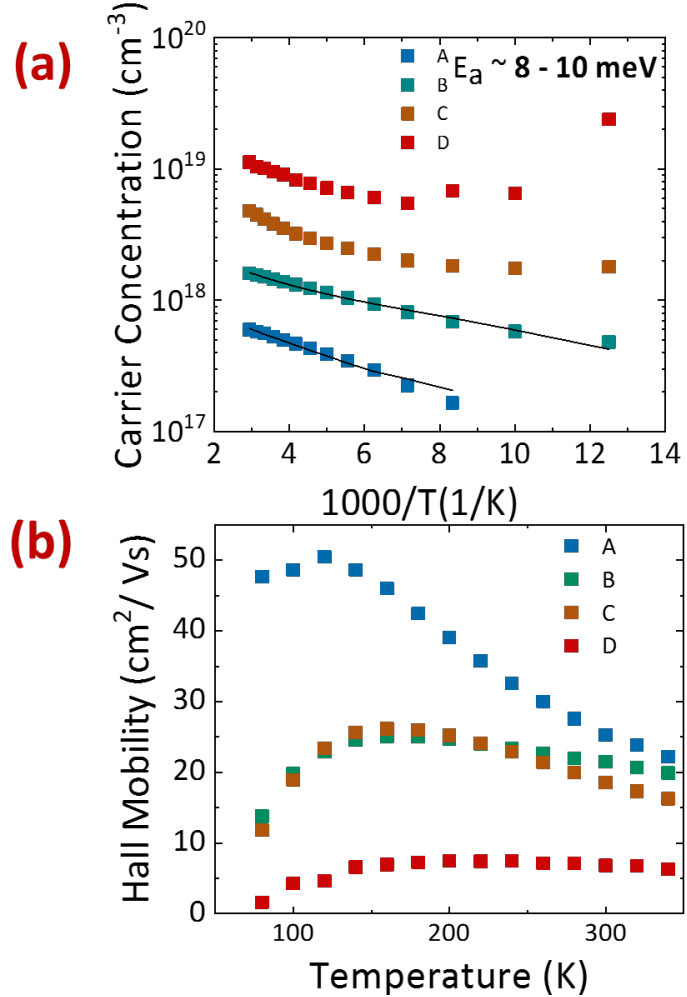

Fig.2 (a) Hall carrier concentration plotted against measurement temperature(1000/T) of Ge-doped β-Ga$_2$O$_3$ grown using setup 1(sample A-D). (b) Corresponding temperature-dependent Hall mobility plotted as a function of temperature.

In order to understand the effect of reactor topology on β-Ga$_2$O$_3$ thin film properties, β-Ga$_2$O$_3$ growths were also performed using setup 2. Using similar parameters, the maximum β-Ga$_2$O$_3$ growth rate increased sharply from <5 μms/hr to up to 22 μms/hr. This is attributed to the increased gallium transport achieved by using the argon delivery tube (setup 2). In this study, growth of β-Ga$_2$O$_3$ is performed on both 6° offcut sapphire wafers (Sample H) and (010) single crystal wafers (Samples E,F,G) from Kyma technologies. Room temperature Hall measurements show that the carrier concentration increased with increasing Ge source temperature, indicating that Ge is the main source of n-type conductivity. A room temperature Hall charge density of 2 x $10^{17}$ – 3 x $10^{18}$ cm$^{-3}$ and Hall mobility of 20 - 35 cm$^2$/Vs were realized. Temperature-dependent Hall measurements were utilized to characterize the dopant density, carrier mobility and activation energies. The low-temperature Hall data is fit to Eq.1, from which the donor densities and activation energies ($N_a, N_{d1}, N_{d2}, E_{d1}, E_{d2}$) are extracted for all the films. The details of the extracted values are listed in Table.1.

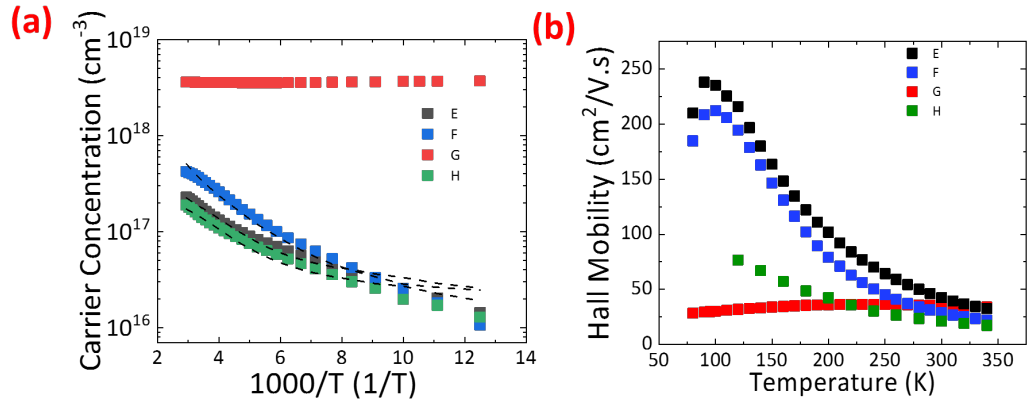

Fig.3 (a) Hall carrier concentration plotted against measurement temperature of Ge-doped β-Ga$_2$O$_3$ grown using setup 2 (sample E-H). Dash line are fits to the charge neutrality equation. (b) Corresponding low-temperature Hall mobility plotted as a function of temperature.

| Sample Name | Sample RT Hall concentration (cm$^{-3}$) | Ea$_1$ (meV) | N$_{d1}$ (cm$^{-3}$) | E$_{d2}$ (meV) | N$_{d2}$ (cm$^{-3}$) | Compensating N$_a$ (cm$^{-3}$) |
|---|---|---|---|---|---|---|
| E | 1.9 x 17 cm$^{-3}$ (010) Homoepitaxial | 19.5 | 8.5 x 10$^{16}$ | 85 | 4.9 x 10$^{17}$ | 1.8 x 10$^{16}$ |
| F | 3.6 x 17 cm$^{-3}$ (010) Homoepitaxial | 19.5 | 1.8 x 10$^{17}$ | 85 | 2 x 10$^{18}$ | 8 x 10$^{16}$ |
| H | 1.5 x 17 cm$^{-3}$ (-201) 6° offcut sapphire | 20 | 6 x 10$^{16}$ | 85 | 3.2 x 10$^{17}$ | 1.1 x 10$^{16}$ |

Table 1. Hall fit parameters of β-Ga$_2$O$_3$ epitaxial films grown using setup 2

Unlike the films grown using setup 1, samples E, F and G showed relatively larger values of mobility at ~100 K. This is attributed to improved material quality due to the homoepitaxial nature of the growth. However, room temperature mobility is still relatively low compared to MBE-grown lightly Ge-doped β-Ga$_2$O$_3$[31]. Sample G, which has a carrier density of 3 x 10$^{18}$ cm$^{-3}$, did not show any change with measurement temperature indicating degenerate behavior. This observation agrees well with other reports of heavily doped β-Ga$_2$O$_3$ with doping greater than 3 x 10$^{18}$ cm$^{-3}$. Low-temperature fits of the carrier concentration profile showed two different donor energy levels, a shallow $E_{d1}$ level at ~ 20 meV and a deeper $E_{d2}$ level at 85 meV. Based on literature reports, the density of the shallow energy level is in general much higher compared to the deeper level ($N_{d1}>N_{d2}, E_{a1}<E_{a2}$). However, in this case the $N_{d2}>N_{d1}$ for all of the three non-degenerate samples(E,F and H). This effect is also independent of sample orientation ((010) or (-201) orientations). However, the present deeper donor level is not seen in samples grown using setup 1. The activation energy plots showef typical carrier freezeout, with no need for using a second donor in the charge neutrality fit. On the other hand, in the case of setup 2, carrier concentration fits could not be obtained without assuming a large density of $E_{d2}$. In the case of sample F, the concentration of $N_{d2}$ is 10x greater than that of $N_{d1}$. This indicates that Ge predominantly occupies $E_{d2}$ level in films grown using setup 2. Based on the above analysis, it is clear that the density of $N_{d2}>N_{d1}$. Additionally, the energy levels ($E_{d1}, E_{d2}$) did not change significantly between various samples. However, the exact origin of the $N_{d1}$ donor levels is not clear. Based on the fitting model, the concentration of $N_{d1}$ is lower than 1 x 10$^{17}$ cm$^{-3}$. This could indicate the presence of an unintentional background donor unrelated to Ge. If that is the case, we would expect $N_{d1}$ to be constant on all the films. However, the concentration of $N_{d1}$ clearly increases with Ge-doping. Also, Hall measurements of UID (un-intentionally doped) material showed no observable n-type conductivity. This observation refutes the background donor hypothesis. We attribute $N_{d2}$ to Ge donors, since such a high concentration of $N_{d2}$ can only be explained through Ge incorporation.

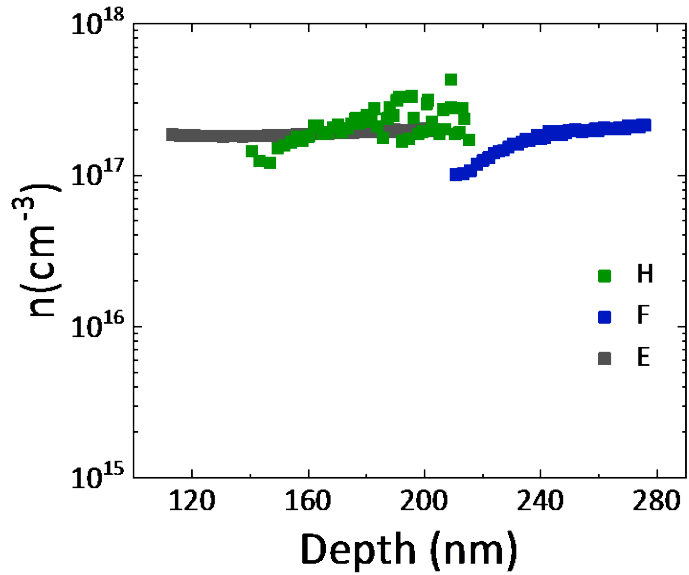

Fig.4 CV extracted apparent carrier density profiles for β-Ga$_2$O$_3$ films grown using setup 2 (Table 1).

CV measurements were done on the as-grown films to confirm the carrier concentration of Ge-doped films. The doping-depth profile of the lateral Schottky diodes in Figure.4. The carrier concentration profiles of samples E,F and H are found close to 2 x10$^{17}$ cm$^{-3}$, which is close to the measured Hall data. Given the thickness of the films and fluctuation in the Ge vapor pressure over time, there could be some doping variation with depth in the film. Nevertheless, the similarity between the CV and Hall data further confirms that Ge is the dominant source of n-type conductivity.

The above data suggests that the reactor configuration has a significant effect on donor incorporation. At low growth rates, Ge forms a shallow donor with a single energy level. When the growth rate increased through higher mass transport of Gallium, Ge prefers to occupy a deeper energy level close to 85 meV. This phenomenon is attributed to a change in dopant incorporation due to surface kinetics. A similar phenomenon has been observed in Si-doped MOVPE grown β-Ga$_2$O$_3$[42]. According to DFT (density functional theory)[43], Ge donor atom can occupy either a tetrahedral or octahedral gallium sites. The calculated donor formation energy in both sites are found to be very close to each other. We hypothesize that the difference in donor energy levels between the two configurations could be explained by the difference in Ge site occupation. Depending on the growth conditions and reactor configuration Ge seems to prefer either tetrahedral or octahedral sites, leading to difference in donor energy levels. However, additional direct experimental studies are required to identify Ge donor occupation as a function of growth condition.

# IV. SUMMARY AND CONCLUSIONS

In summary, we developed a new technique for achieving n-type doping in LPCVD-grown β-Ga$_2$O$_3$ thin films using solid source dopants. LPCVD growth of Ge-doped β-Ga$_2$O$_3$ thin films is performed using solid-source germanium as dopant source. LPCVD growth of Ge-doped β-Ga$_2$O$_3$ was studied using two different reactor configurations (Gallium and Oxygen delivery tubes). By controlling the Ge vapor pressure, carrier concentration between 1 x 10$^{17}$ cm$^{-3}$ and 1 x 10$^{19}$ cm$^{-3}$ was attained for films on 6˚ degree offcut and (010) β-Ga$_2$O$_3$. At low growth rates, Ge prefers to occupy a shallow donor level with E$_d$ of 8 – 10 meV. When the growth rate increased to ~ 22 μm/hr, Ge is prefers to occupy a shallow level N$_{d1}$ (20 meV) and a deep level N$_{d2}$ (85 meV). The density of the deeper level was found to be much larger than the shallow level N$_{d1}$. This study shows the effect of growth kinetics on Ge incorporation. Studying such phenomena is important for the design of β-Ga$_2$O$_3$ based high-power electronic devices.

## ACKNOWLEDGMENTS

This material is based upon work supported by the Air Force Office of Scientific Research under award number FA9550-18-1-0507 and monitored by Dr. Ali Sayir. Any opinions, findings, conclusions, or recommendations expressed in this material are those of the authors and do not necessarily reflect the views of the United States Air Force. Praneeth Ranga acknowledges support from University of Utah Graduate Research Fellowship 2020-2021. This work was performed in part at the Utah Nanofab sponsored by the College of Engineering and the Office of the Vice President for Research. We also thank Jonathan Ogle for his support regarding Hall measurements.

## DATA AVAILABILITY

The data that support the findings of this study are available from the corresponding author upon reasonable request.

# REFERENCES

[1] S.J. Pearton, F. Ren, M. Tadjer, and J. Kim, J. Appl. Phys. **124**, 220901 (2018).

[2] M. Higashiwaki, K. Sasaki, A. Kuramata, T. Masui, and S. Yamakoshi, Appl. Phys. Lett. **100**, 013504 (2012).

[3] A. Kuramata, K. Koshi, S. Watanabe, Y. Yamaoka, T. Masui, and S. Yamakoshi, Jpn. J. Appl. Phys. **55**, 1202A2 (2016).

[4] K. Goto, K. Konishi, H. Murakami, Y. Kumagai, B. Monemar, M. Higashiwaki, A. Kuramata, and S. Yamakoshi, Thin Solid Films **666**, 182 (2018).

[5] Y. Zhang, F. Alema, A. Mauze, O.S. Koksaldi, R. Miller, A. Osinsky, and J.S. Speck, APL Mater. **7**, 022506 (2019).

[6] A. Bhattacharyya, P. Ranga, S. Roy, J. Ogle, L. Whittaker-Brooks, and S. Krishnamoorthy, Appl. Phys. Lett. **117**, 142102 (2020).

[7] Z. Feng, A.F.M.A.U. Bhuiyan, Z. Xia, W. Moore, Z. Chen, J.F. McGlone, D.R. Daughton, A.R. Arehart, S.A. Ringel, S. Rajan, and H. Zhao, Phys. Status Solidi RRL – Rapid Res. Lett. **14**, 2000145 (2020).

[8] S. Sharma, K. Zeng, S. Saha, and U. Singisetti, IEEE Electron Device Lett. **41**, 836 (2020).

[9] W. Li, Z. Hu, K. Nomoto, R. Jinno, Z. Zhang, T.Q. Tu, K. Sasaki, A. Kuramata, D. Jena, and H.G. Xing, in *2018 IEEE Int. Electron Devices Meet. IEDM* (2018), p. 8.5.1-8.5.4.

[10] K. Ghosh and U. Singisetti, Appl. Phys. Lett. **109**, 072102 (2016).

[11] A. Kumar, K. Ghosh, and U. Singisetti, J. Appl. Phys. **128**, 105703 (2020).

[12] A.J. Green, K.D. Chabak, E.R. Heller, R.C. Fitch, M. Baldini, A. Fiedler, K. Irmscher, G. Wagner, Z. Galazka, S.E. Tetlak, A. Crespo, K. Leedy, and G.H. Jessen, IEEE Electron Device Lett. **37**, 902 (2016).

[13] Z. Xia, H. Chandrasekar, W. Moore, C. Wang, A.J. Lee, J. McGlone, N.K. Kalarickal, A. Arehart, S. Ringel, F. Yang, and S. Rajan, Appl. Phys. Lett. **115**, 252104 (2019).

[14] W. Li, K. Nomoto, Z. Hu, D. Jena, and H.G. Xing, IEEE Electron Device Lett. **41**, 107 (2020).

[15] Y. Zhang, A. Neal, Z. Xia, C. Joishi, J.M. Johnson, Y. Zheng, S. Bajaj, M. Brenner, D. Dorsey, K. Chabak, G. Jessen, J. Hwang, S. Mou, J.P. Heremans, and S. Rajan, Appl. Phys. Lett. **112**, 173502 (2018).

[16] P. Ranga, A. Bhattacharyya, A. Chmielewski, S. Roy, R. Sun, M.A. Scarpulla, N. Alem, and S. Krishnamoorthy, Appl. Phys. Express **14**, 025501 (2021).

[17] P. Ranga, A. Bhattacharyya, A. Rishinaramangalam, Y.K. Ooi, M.A. Scarpulla, D. Feezell, and S. Krishnamoorthy, Appl. Phys. Express **13**, 045501 (2020).

[18] N.K. Kalarickal, Z. Xia, J.F. McGlone, Y. Liu, W. Moore, A.R. Arehart, S.A. Ringel, and S. Rajan, J. Appl. Phys. **127**, 215706 (2020).

[19] Y. Lv, Y. Wang, X. Fu, S. Dun, Z. Sun, H. Liu, X. Zhou, X. Song, K. Dang, S. Liang, J. Zhang, H. Zhou, Z. Feng, S. Cai, and Y. Hao, IEEE Trans. Power Electron. **36**, 6179 (2021).

[20] A.T. Neal, S. Mou, R. Lopez, J.V. Li, D.B. Thomson, K.D. Chabak, and G.H. Jessen, Sci. Rep. **7**, (2017).

[21] K. Sasaki, A. Kuramata, T. Masui, E.G. Víllora, K. Shimamura, and S. Yamakoshi, Appl. Phys. Express **5**, 035502 (2012).

[22] Z. Feng, A.F.M. Anhar Uddin Bhuiyan, M.R. Karim, and H. Zhao, Appl. Phys. Lett. **114**, 250601 (2019).

[23] S. Bin Anooz, R. Grüneberg, C. Wouters, R. Schewski, M. Albrecht, A. Fiedler, K. Irmscher, Z. Galazka, W. Miller, G. Wagner, J. Schwarzkopf, and A. Popp, Appl. Phys. Lett. **116**, 182106 (2020).

[24] J.H. Leach, K. Udwary, J. Rumsey, G. Dodson, H. Splawn, and K.R. Evans, APL Mater. **7**, 022504 (2019).

[25] K.D. Leedy, K.D. Chabak, V. Vasilyev, D.C. Look, J.J. Boeckl, J.L. Brown, S.E. Tetlak, A.J. Green, N.A. Moser, A. Crespo, D.B. Thomson, R.C. Fitch, J.P. McCandless, and G.H. Jessen, Appl. Phys. Lett. **111**, 012103 (2017).

[26] Z. Feng, M.R. Karim, and H. Zhao, APL Mater. **7**, 022514 (2019).

[27] S. Rafique, L. Han, A.T. Neal, S. Mou, J. Boeckl, and H. Zhao, Phys. Status Solidi A **215**, 1700467 (2018).

[28] S. Rafique, L. Han, A.T. Neal, S. Mou, M.J. Tadjer, R.H. French, and H. Zhao, Appl. Phys. Lett. **109**, 132103 (2016).

[29] G. Joshi, Y.S. Chauhan, and A. Verma, ArXiv Prepr. ArXiv210203833 (2021).

[30] Y. Oshima, E. Ahmadi, S. Kaun, F. Wu, and J.S. Speck, Semicond. Sci. Technol. **33**, 015013 (2018).

[31] E. Ahmadi, O.S. Koksaldi, S.W. Kaun, Y. Oshima, D.B. Short, U.K. Mishra, and J.S. Speck, Appl. Phys. Express **10**, 041102 (2017).

[32] T. Itoh, A. Mauze, Y. Zhang, and J.S. Speck, Appl. Phys. Lett. **117**, 152105 (2020).

[33] A. Mauze, Y. Zhang, T. Mates, F. Wu, and J.S. Speck, Appl. Phys. Lett. **115**, 052102 (2019).

[34] N.K. Kalarickal, Z. Xia, J. McGlone, S. Krishnamoorthy, W. Moore, M. Brenner, A.R. Arehart, S.A. Ringel, and S. Rajan, Appl. Phys. Lett. **115**, 152106 (2019).

[35] A. Hassa, H. von Wenckstern, D. Splith, C. Sturm, M. Kneiß, V. Prozheeva, and M. Grundmann, APL Mater. **7**, 022525 (2019).

[36] H. Murakami, K. Nomura, K. Goto, K. Sasaki, K. Kawara, Q.T. Thieu, R. Togashi, Y. Kumagai, M. Higashiwaki, A. Kuramata, S. Yamakoshi, B. Monemar, and A. Koukitu, Appl. Phys. Express **8**, 015503 (2015).

[37] Y. Oshima, E.G. Víllora, and K. Shimamura, J. Cryst. Growth **410**, 53 (2015).

[38] S. Rafique, M.R. Karim, J.M. Johnson, J. Hwang, and H. Zhao, Appl. Phys. Lett. **112**, 052104 (2018).

[39] Y. Zhang, Z. Feng, M.R. Karim, and H. Zhao, J. Vac. Sci. Technol. A **38**, 050806 (2020).

[40] C. Joishi, S. Rafique, Z. Xia, L. Han, S. Krishnamoorthy, Y. Zhang, S. Lodha, H. Zhao, and S. Rajan, Appl. Phys. Express **11**, 031101 (2018).

[41] Z. Kabilova, C. Kurdak, and R.L. Peterson, Semicond. Sci. Technol. **34**, 03LT02 (2019).

[42] A. Fiedler, "Electrical and optical characterization of b-Ga2O3," Ph.D. thesis (Humboldt University at Berlin, 2019).

[43] J. B. Varley, First-principles calculations 2 - doping and defects in Ga2O3, in Gallium Oxide - Materials Properties, Crystal Growth, and Devices, edited by M. Higashiwaki and S. Fujita (Springer, 2020)